\documentclass[twocolumn,twoside,slac_two]{revtex4}
\usepackage{graphicx}
\usepackage{fancyhdr}
\newcommand{\fermi}{\emph{Fermi}}
\newcommand{\DG}{$^{\circ}$}
\newcommand{\al}{$\alpha$}
\newcommand{\ze}{$\zeta$}
\newcommand{\fdg}{\mbox{\ensuremath{.\!\!^\circ}}}

\begin{document}

\title{Modeling and Maximum Likelihood Fitting of Gamma-Ray and Radio Light Curves of Millisecond Pulsars Detected with \emph{Fermi}}

\author{T.~J.~Johnson}
\affiliation{Departments of Physics and Astronomy, University of Maryland, College Park, MD 20742, USA: to whom correspondence should be addressed, tyrel.j.johnson@gmail.com}
\author{A.~K.~Harding}
\affiliation{Astrophysics Science Division, NASA Goddard Space Flight Center, Greenbelt, MD 20771, USA}
\author{C.~Venter}
\affiliation{Centre for Space Research, North-West University, Potchefstroom Campus, Private Bag X6001, Potchefstroom 2520, South Africa}

\begin{abstract}
Pulsed gamma rays have been detected with the \fermi{} Large Area Telescope (LAT) from more than 20 millisecond pulsars (MSPs), some of which were discovered in radio observations of bright, unassociated LAT sources. We have fit the radio and gamma-ray light curves of 19 LAT-detected MSPs in the context of geometric, outer-magnetospheric emission models assuming the retarded vacuum dipole magnetic field using a Markov chain Monte Carlo maximum likelihood technique. We find that, in many cases, the models are able to reproduce the observed light curves well and provide constraints on the viewing geometries that are in agreement with those from radio polarization measurements. Additionally, for some MSPs we constrain the altitudes of both the gamma-ray and radio emission regions.  The best-fit magnetic inclination angles are found to cover a broader range than those of non-recycled gamma-ray pulsars.
\end{abstract}

\maketitle

\thispagestyle{fancy}

\section{INTRODUCTION}\label{intro}
Millisecond pulsars (MSPs) are thought to be old, recycled objects which have reached short spin periods ($\lesssim$ 10 ms) via accretion from a binary companion \cite{Alpar82}.  MSPs have been established as a population of high-energy (HE, $\geq$ 0.1 GeV) emitters via the detection of significant pulsed gamma-ray signals from more than 20 MSPs at the radio periods using data from the \fermi{} Large Area Telescope (LAT).  Additionally, more than 30 previously unknown MSPs have been found in radio observations of unassociated LAT sources, some of which have already been confirmed as gamma-ray pulsars \cite{Ransom11,Cognard11,Keith11}.  MSPs can be sorted into three sub-classes: those for which the gamma-ray peaks lag the main radio component (class I), those for which the gamma-ray and radio profile components are aligned in phase (class II), and those which lead the main radio component (class III).  Non-recycled gamma-ray pulsars do not show such a diverse distribution of light curves; thus, studies of MSP gamma-ray and radio light curves provide a much broader window into the pulsar magnetosphere and emission mechanisms.  We have generated simulations of different emission models assuming a vacuum retarded dipole magnetic field geometry \cite{Deutsch55} and attempted to reproduce the gamma-ray and radio light curves of 19 MSPs detected with the \fermi{} LAT using 2 years of data and a maximum likelihood fitting technique.

\section{GAMMA-RAY EMISSION MODELS}\label{gmodels}
To reproduce the gamma-ray light curves of MSPs in class I we have used two-pole caustic (TPC; \cite{DR03}) and outer gap (OG; e.g., \cite{Cheng86}) models.  Note that we take the TPC model to be a geometric realization of the slot gap (SG) model \cite{MH04} and thus some of the parameters of our TPC model differ from those of \cite{DR03} as discussed in Section \ref{sims}.

The TPC/SG model assumes that particles are accelerated and HE gamma rays are emitted in narrow gaps along the surface of last-closed field lines.  The gaps start at the stellar surface and continue out to the light cylinder (defined by the cylindrical radius R$_{\rm LC}\ =\ c/\Omega$, where $\Omega$ is the rotational frequency of the pulsar).  The OG model assumes that a vacuum accelerating gap forms along the surface of last-closed field lines but only above the null-charge surface (NCS, defined by the requirement that $\vec{\Omega}\cdot\vec{B}\ =\ 0$).  In the OG model, HE gamma rays are emitted in a thin layer interior to the accelerating gap.

In both models the bulk of the HE emission originates at high altitudes above the stellar surface, near the light cylinder.  The bright, sharp peaks observed in HE pulsar light curves \cite{PSRcat} are thought to be emission caustics \cite{Morini83} where relativistic aberration, time-of-flight delays, and magnetic field line curvature combine to cancel out phase differences between photons emitted at different altitudes.  This results in many photons arriving at the observer at nearly the same phase.

The TPC and OG models have also been used to reproduce the gamma-ray light curves of those MSPs in class II; however, given the alignment in phase of the gamma-ray and radio peaks for these pulsars we fit for the maximum radial extent to which the gamma-ray emission is followed, see Sections \ref{sims} and \ref{fit} for more details.  We identify these models as altitude-limited TPC and OG (alTPC and alOG, respectively) in order to distinguish them from the models used for MSPs in class I, for more details on the alTPC/OG models see \cite{Venter11}.

The periods (P) of MSPs are found to be very stable with small period derivatives ($\dot{\rm P} \lesssim 10^{-17}\ \rm s\ s^{-1}$).  Such relatively low spin-down rates place the majority of MSPs below the pair-creation death line on a $\dot{\rm P}$ vs. P plot \cite{HM02}.  Pulsars below this death line were not expected to be capable of screening the accelerating electric field over most of the open volume and thus creating narrow accelerating gaps required in SG and OG models.  Thus, \cite{Harding05} created the pair-starved polar cap (PSPC) model for HE gamma-ray emission in pulsars below the death line.  The PSPC model has been used to reproduce the gamma-ray light curves of MSPs in class III.  In this model the entire open volume (from the magnetic dipole axis to the surface of last-closed field lines) is available to accelerate particles which emit HE gamma rays.

\section{RADIO EMISSION MODELS}\label{rmodels}
To reproduce the radio light curves of MSPs in classes I and III we have used a single-height, hollow-cone beam following the description of \cite{Story07} and \cite{Harding08}.  This model assumes that the radio emission originates at an altitude which depends on P, the frequency of emission, and weakly on $\dot{\rm P}$ \cite{KG03}.  For typical MSP periods and observing frequencies this altitude is $\lesssim\ 30 \%\ \rm R_{LC}$.  This is significantly lower in altitude than the bulk of the HE emission and thus leads to the non-zero phase offsets between radio and gamma-ray light curve features for MSPs in classes I and III.

Radio and gamma-ray light curves with features which are aligned in phase imply, at least partial, co-location of the emission regions.  Thus, to reproduce the radio light curves of MSPs in class II we used alTPC/OG models in which we fit both the minimum and maximum radio emission altitudes.  This implies that the radio emission is also caustic in nature which has important implications for the predicted polarization properties \cite{Venter11}.

\section{LIGHT CURVE SIMULATIONS}\label{sims}
We have generated simulations with spin periods of 1.5, 2.5, 3.5, 4.5, and 5.5 ms for MSPs in classes I and III.  For MSPs of class II we have only generated simulations with a spin period of 1.5 ms.  This is shorter than the observed periods of PSRs J2214+3000 \cite{Ransom11} and J1823$-$3021A \cite{Freire11} but will, at most, slightly overestimate any predicted off-pulse interval.

All of our simulations have a resolution of 2\DG{} in pulse phase and 1\DG{} in both magnetic inclination angle (\al) and observer viewing angle (\ze).  For MSPs in class I our simulations have a resolution of 5\% of the polar cap opening angle for the accelerating and emitting gap widths.  For MSPs in class II our simulations have the same resolution in gap widths and a resolution of 0.1R$_{\rm LC}$ in emission altitude.

Our simulation code follows that of \cite{DHR04} with a few important modifications.  We have included the Lorentz transformation of the magnetic field from an inertial observer's frame to the frame which co-rotates with the star before calculating direction tangent to the field line along which a photon is emitted (see Appendix B of \cite{Johnson11} for more details) as advocated by \cite{BS10} for self-consistency.  Additionally, for the PSPC models we have used the same functional form for the accelerating field as \cite{Venter09} to calculate the number of photons emitted at each step along the field lines as opposed to assuming uniform emissivity as is done for the other models.  For all models the emission is never followed beyond a radial distance of 1.2R$_{\rm LC}$ or a cylindrical distance of 0.95R$_{\rm LC}$, whichever is reached first.  In this respect our TPC models differ from those originally used by \cite{DR03} as they only followed emission out to a radial distance of 0.95R$_{\rm LC}$ but not beyond a cylindrical distance of 0.75R$_{\rm LC}$.

\section{LIKELIHOOD FITTING}\label{fit}
We have developed a Markov chain Monte Carlo (MCMC) maximum likelihood technique in order to pick the best-fit model parameters.  An MCMC procedure involves taking random steps in parameter space and accepting or rejecting new steps based on some criteria which only involves the previous step (in our case it is the likelihood ratio between the two states).  In order to speed up chain convergence and mixing we have implemented small-world chain steps \cite{Guan06} and simulated annealing \cite{MP92}.  We use Poisson likelihood for the gamma-ray light curves and a $\chi^{2}$ statistic for the radio, combine them, and then maximize the joint likelihood.

The formal uncertainty on the radio profiles is much smaller, relatively, than that of the gamma-ray light curves which drives the likelihood to favor the radio fit.  In order to balance the relative contributions to the likelihood from the radio and gamma-ray data we have chosen to use the same uncertainty for each bin of the radio light curve which is defined as follows.  We first calculate the average relative error in the on-peak interval of the gamma-ray light curve.  Then, we multiply that value with the maximum of the radio profile.  Finally, in the event that we use more bins in the radio profile than the gamma-ray light curve we multiply the latter uncertainty by the ratio of gamma-ray to radio bins.  It is important to note that using a different radio uncertainty can affect the best-fit geometry, in some cases leading to changes in either \al{} or \ze{} of $\sim$30\DG{}.

\section{RESULTS}\label{res}
\begin{figure}[h]
\includegraphics[width=85mm]{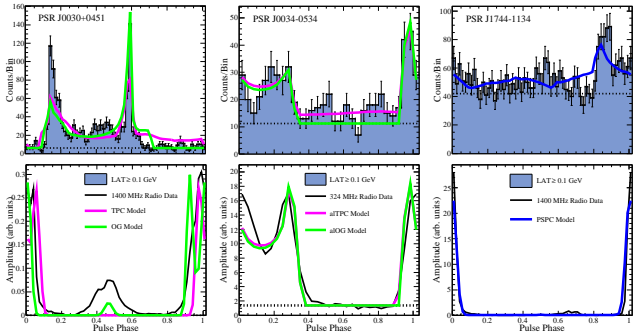}
\caption{Example light curve fits representing the three different MSP sub-classes.  Gamma-ray data and model light curves are shown in the top panels while the radio data and model light curves are shown in the lower panels.  \emph{(Left:)} Light curves for PSR J0030+0451 fit with OG and TPC models (class I).  \emph{(Middle:)} Light curves for PSR J0034$-$0534 fit with the alOG and alTPC models (class II).  \emph{(Right:)} Light curves for PSR J1744$-$1134 fit with the PSPC model (class III).\label{example}}
\end{figure}

Example fits from each MSP sub-class are shown in Fig.~\ref{example}.  The gamma-ray events for each MSP were required to be within $0\fdg8$ of the radio position, have reconstructed energies from 0.1 to 100 GeV, and have zenith angles $\leq$ 105\DG{}.  The background levels for the gamma-ray light curves were estimated using the LAT Science Tool \texttt{gtsrcprob} and spectral results from a preliminary version of the 2FGL catalog\footnote{http://fermi.gsfc.nasa.gov/ssc/data/access/\\lat/2yr\_catalog/} \cite{2FGL} while the radio backgrounds are estimated by fitting a constant value to the off-pulse intervals.  In Fig.~\ref{example} the model light curves corresponding to TPC and alTPC fits are shown in pink, the OG and alOG in green, and the PSPC in blue.

\begin{figure}[h]
\includegraphics[width=85mm]{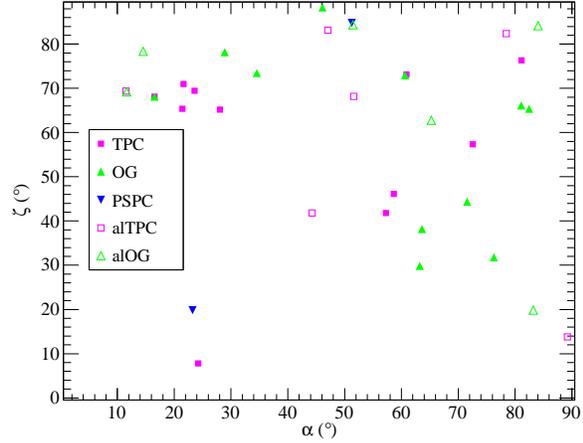}
\caption{Best-fit (\ze,\al) values for each MSP (both fits are shown for MSPs in classes I and II).  TPC fit values are shown as filled pink squares, OG as filled green triangles, and PSPC as filled blue triangles.  The alTPC fits are shown as open pink squares and the alOG as open green triangles.\label{geom}}
\end{figure}

When plotting the best-fit geometries (Fig.~\ref{geom}) an interesting trend appears.  The \ze{} values tend to prefer higher angles near 90\DG{}; this is consistent with the assumption of a random angular distribution (i.e., weighted by $\sin(\zeta)$) of spin axes with respect to the Earth line-of-sight.  However, the \al{} values seem to favor all angles equally in contrast to what has been found for non-recycled gamma-ray pulsars \cite{Pierbattista11}.  This may be a manifestation of the recycling process tending to align the magnetic and spin axes.

\begin{figure}[h]
\includegraphics[width=85mm]{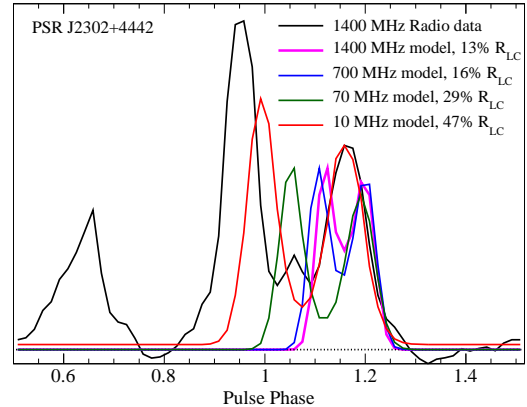}
\caption{Radio data and models with different emission altitudes for PSR J2302+4442.\label{radioup}}
\end{figure}

There are some suggestions in our fits that the assumed radio emission altitude of the conal model (classes I and III) is too low.  One such example is PSR J2302+4442 for which the gamma-ray model reproduces the observed light curve well but the radio model can not produce widely enough separated peaks at the best-fit geometry \cite{Cognard11}.  Fig.~\ref{radioup} demonstrates how, for the same geometry which gives a good gamma-ray fit, increasing the radio emission altitude does, in fact, lead to a better fit to the radio data. Note that the radio altitude was increased in the simulations by decreasing the emission frequency.

\section{FUTURE}\label{future}
The simulations described in Section \ref{sims} have rather coarse resolution in the gap width and altitude parameters and we have found that a finer gridding is needed before we can look for meaningful trends in these values.  In some of the MSPs we have analyzed the single, hollow-cone beam model is not correct as there is evidence for either a core component or multiple cones from radio polarization data.  Additionally, as noted in Section \ref{res}, we find indications that the radio emission should originate at higher altitudes in the MSP magnetospheres.  Therefore, in addition to increasing the resolution of our simulations we plan to produce more complex radio models and explore changing the emission altitude in order to more closely match the radio profiles.  We will also compare the predicted polarization angle swings with polarimetric data to guide further model refinement and serve as an additional, observational test.

The magnetosphere of a pulsar should be filled, to some extent, with charges (e.g., \cite{GJ69}); thus, the vacuum solution for the magnetic field can not exactly match reality.  With that in mind, we plan to apply the same fitting technique to simulations using magnetic field geometries from magneto-hydrodynamic simulations of a pulsar magnetosphere under force-free assumptions (e.g., \cite{Contopolous99}) and with finite conductivity \cite{Kalapotharakos11}.  By comparing our predicted gamma-ray and radio light curves in different magnetic field geometries with those from observations, we can constrain the structure of a real pulsar magnetosphere.

\begin{acknowledgments}
This research was supported by a \fermi{} Cycle III GI grant.  This research was performed while TJJ was a graduate student at the University of Maryland but he is now a National Research Council Research Associate at the Naval Research Laboratory and is sponsored by NASA DPR S-15633-Y.   CV acknowledges support from the South African National Research Foundation.  AKH acknowledges support from the NASA Astrophysics Theory Program.
\end{acknowledgments}

\end{document}